\documentclass[a4paper]{article}

\title {Quantum axiomatics and a theorem of M.P.~Sol\`{e}r\footnote{Published as: Aerts, D. and Van
Steirteghem, B., 2000, ``Quantum axiomatics and a theorem of M.P.~Sol\`{e}r", {\it International Journal of
Theoretical Physics}, {\bf 39}, 497-502.}}
\author {Diederik Aerts, Bart Van Steirteghem\thanks{FUND, Department
of Mathematics, Brussels Free University, Pleinlaan 2, B--1050 Brussels;
e-mails: diraerts@vub.ac.be, bvsteirt@vub.ac.be}}

\date {}

\newtheorem{proposition}{Proposition}

\newcommand{\inn}{\subset}

\newcommand{\sep}{\hskip -2pt \makebox[11pt]{$\bigcirc$ \hskip -11.5 pt
\raise 0.1 pt \hbox{$\wedge$}}}

\newcommand{\smallsep}{\footnotesize\hskip -2pt \makebox[11pt]{$\bigcirc$
\hskip -10 pt \raise 0.1 pt \hbox{$\wedge$}}}

\usepackage{indentfirst}
\usepackage[psamsfonts]{amssymb}
\newcommand{\compl}{{\mathbb C}}
\newcommand{\real}{{\mathbb R}}

\newcommand{\quat}{{\mathbb H}}

\begin{document}

\maketitle

\begin{abstract}
\noindent Three of the traditional quantum axioms (orthocomplementation,
orthomodularity and the covering law) show incompatibilities with two
products introduced by Aerts for the description of joint entities. Inspired
by Sol\`er's theorem and Holland's AUG axiom, we propose a property of `plane
transitivity', which also characterizes classical Hilbert spaces among
infinite--dimensional orthomodular spaces, as a possible partial substitute
for the `defective' axioms.
\end{abstract}

\section{Introduction}
\noindent 
In his axiomatization of standard quantum mechanics
Holland (1995) introduces the Ample Unitary Group axiom (cf.\ (2) in
Proposition~\ref{prop1} of this paper). It hints at an evolution axiom but
has the shortcoming that it is not lattice theoretical. In particular, it
cannot be formulated for \emph{property lattices} ---complete, atomistic
and orthocomplemented
lattices--- which play a central role in the Geneva--Brussels approach to
the foundations of
physics (Piron, 1976, 1989,
1990; Aerts, 1982, 1983, 1984; Moore, 1995, 1999). Inspired by this axiom, we
propose a
property, called `plane transitivity' (section~\ref{secpt}), which does not
have
{\em this} imperfection. Like the AUG axiom it characterizes classical Hilbert
spaces among infinite--dimensional orthomodular spaces and it still looks like
`demanding enough symmetries or evolutions'.

The traditional quantum axiomatics show some shortcomings in
the description of compound systems (Aerts, 1982,
1984; Pulmannov\`a, 1983, 1985). In particular, orthocomplementation,
orthomodularity and the covering law are not compatible with two products
---`separated' and `minimal'--- introduced by Aerts (section~\ref{secce}).
Plane transitivity, on the other hand, `survives' these two products
(section~\ref{secpt}). That way it is a candidate to help
fill the gap left by the failing axioms.

\section{Alternatives to Sol\`er's theorem}
\noindent
Consider a complete, atomistic, orthocomplemented and irreducible
lattice $\cal L$ satisfying the covering law. Suppose moreover that its
length is at least
$4$. Then there exist a division ring $K$ with an involutorial
anti--automorphism $\lambda \mapsto \lambda^*$ and a vector
space $E$ over $K$ with a Hermitian form $<\cdot,\cdot >$ such that
$\cal L$ is ortho--isomorphic to the lattice ${\cal L}(E)$ of closed
(biorthogonal) subspaces of $E$. Moreover, $\cal L$ is orthomodular if and
only if
$(E,K,<\cdot,\cdot>)$ is orthomodular: $M+M^{\perp}=E$ for every
$\emptyset \not=M \subset E$ with $M=M^{\perp\perp}$ (Maeda and Maeda,
1970; Piron, 1976; Faure and Fr\"olicher, 1995). Sol\`er (1995) has proven
the following characterization of classical Hilbert spaces: if
$E$ contains an infinite orthonormal sequence, then $K = \real, \compl$ or
$\quat$ and $(E,K,<\cdot,\cdot>)$ is the corresponding Hilbert space.
Holland (1995) has shown that it is enough to demand the existence of a nonzero
$\lambda \in K$ and an infinite orthogonal sequence $(e_n)_n \in E$ such that
$<e_n,e_n>=\lambda$ for every $n$. To be precise, either
$(E,K,<\cdot,\cdot>)$ or
$(E,K,-<\cdot,\cdot>)$ is then a classical Hilbert space. We shall not make
this precision
explicitly in what follows.

In the first proposition, we summarize some alternatives to Sol\`er's
result, by means of automorphisms of ${\cal L}(E)$.

\begin{proposition} \label{prop1}
Let $(E,K,<\cdot,\cdot>)$ be an orthomodular space and let ${\cal L}(E)$ be
the lattice of its closed subspaces. The following are equivalent:

(1) $(E,K,<\cdot,\cdot>)$ is an infinite--dimensional Hilbert space over $K =
\real,
\compl$ or
$\quat$.

(2) $E$ is infinite--dimensional and given two orthogonal atoms
$p,q$ in ${\cal L}(E)$, there is a unitary operator $U$ such that $U(p)=q$.

(3) There exist $a,b \in {\cal L}(E)$, where $b$ is of dimension at least
$2$, and an ortholattice automorphism $f$ of ${\cal L}(E)$ such that
$f(a)\lneq a$ and $f|_{[0,b]}$ is the identical map.

(4) $E$ is infinite--dimensional and given two orthogonal atoms $p,q$ in
${\cal L}(E)$ there exist distinct atoms $p_1, p_2$ and an
ortholattice automorphism $f$ of ${\cal L}(E)$ such that $f|_{[0,p_1\vee
p_2]}$ is the identity and $f(p)=q$.
\end{proposition}
Condition (2) is Holland's Ample Unitary Group axiom (1995) and (3) is due to
Mayet (1998). Using the properties listed in section 2 of (Mayet, 1998),
one can easily prove that (4) implies (2). We will use (4) to formulate a
lattice theoretical alternative to the AUG axiom (section~\ref{secpt}).

\section{Compound entities and the axioms}\label{secce}
\noindent
Aerts has introduced two `products' for the description of compound
entities. We shall present them `mathematically' and recall their
`interaction' with the axioms of quantum mechanics proposed by Piron
(1976). For an operational justification of these products we refer to
(Aerts, 1982, 1984).

First we recall some notions and results due to Moore (1995). A \emph{state
space} is a pair $(\Sigma,\perp)$, where $\Sigma$ is a set (of states) and
$\perp$ (orthogonality) is a symmetric antireflexive binary relation which
separates the points of $\Sigma$ (if $p\not=q$ then $\exists r$
such that $p\perp r$ and $q \not\perp r$). For $A \subset \Sigma$, put
$A^{\perp}=\{q\in\Sigma\, |\,q\perp p\ \forall p \in A\}$. Then $({\cal
L}_{\Sigma},\subset,\, ^{\perp})$ is a property lattice with
$\bigwedge\{A_r\}=\bigcap\{A_r\}$, where ${\cal L}_{\Sigma}=\{A \subset
\Sigma\, |\, A=A^{\perp\perp}\}$. In particular, $(\Sigma,{\cal
L}_{\Sigma})$ is a T$_{_1}$--\emph{closure space}: ${\cal
L}_{\Sigma}\ni\emptyset$ is a family of subsets of $\Sigma$ closed under
arbitrary intersections and
$\{p\}\in {\cal L}_{\Sigma}, \forall p\in \Sigma$.

Next, consider two entities $S_1, S_2$ described by their state
spaces
$(\Sigma_1,\perp_1)$ and $(\Sigma_2,\perp_2)$. Denote the corresponding
property lattices by ${\cal L}_1$ and ${\cal L}_2$. Suppose $S_1$ and
$S_2$ are `separated'. Aerts (1982) suggests the {\em separated product}
${\cal L}_1\sep{\cal L}_2$ for the description of $S_1$ and $S_2$ taken
together. Its state space is $(\Sigma_1\times\Sigma_2,\perp_{_{\smallsep}})$
where
$$(p_1,p_2)\perp_{_{\smallsep}}(q_1,q_2) \Leftrightarrow p_1\perp q_1 \mbox{
or } p_2\perp q_2$$ ${\cal L}_1\sep{\cal L}_2$ is then the corresponding
property lattice (Piron, 1989). This product is not `compatible' with
orthomodularity and the covering law in the following sense: if ${\cal
L}_1\sep{\cal L}_2$ satisfies one of these properties, then ${\cal L}_1$
or ${\cal L}_2$ is Boolean (Aerts, 1982).

Aerts (1984) proposes another lattice as the `coarsest'
description of a compound entity containing the two (not necessarily
separated) entities $S_1,S_2$. We give a slightly different, but
equivalent construction of this {\em minimal product} ${\cal L}_1\coprod
{\cal L}_2$. Consider the closure spaces $(\Sigma_i, {\cal L}_i)$.
Since ${\bf Cls}_{_1}$, the category of T$_{_1}$--closure spaces and
continuous maps, is closed under products (cf.\ Dikranjan et al., 1988),
$(\Sigma_1,{\cal L}_1)$ and
$(\Sigma_2,{\cal L}_2)$  have a
${\bf Cls}_{_1}$--product, which we denote $(\Sigma_1\times\Sigma_2, {\cal
L}_1\coprod {\cal L}_2)$. This notation is for consistency with
(Aerts et al., 1999). Of course, ${\cal L}_1\coprod{\cal L}_2$ is a
complete atomistic lattice, but the orthocomplementation is problematic.
Indeed, if we define the following
---operationally justified by Aerts (1984)--- orthogonality on
$\Sigma_1\times\Sigma_2$:
$(p_1,p_2)\perp(q_1,q_2) \Leftrightarrow p_1\perp q_1 \mbox{ or }
p_2\perp q_2$, then ${\cal L}_1\coprod{\cal L}_2$ cannot have an
orthocomplementation compatible with $\perp$ unless ${\cal L}_1 \inn \{0,1\}$
or
${\cal L}_2 \inn \{0,1\}$. Moreover, the same is true for the covering
law: if ${\cal L}_1\coprod{\cal L}_2$ satisfies the covering law, then
${\cal L}_1$ or ${\cal L}_2$ is trivial. For completeness, we mention that
this product is compatible with a suitable form of orthomodularity.

These problems with the traditional axioms in the description of joint
entities have made it desirable to find (nice) properties compatible with
the separated and minimal product. If we slightly generalize condition (4) of
Proposition~\ref{prop1}, we obtain a property which survives
both products.

\section{Plane transitivity} \label{secpt}
\noindent
To seize both products with the same terminology, we introduce \emph{pseudo
property lattices}. $({\cal L},\Sigma,\perp)$ is a {\em p.p.l.}\ if ${\cal
L}$ is a complete atomistic lattice and $\perp$ is an orthogonality on its
set of atoms $\Sigma$. Using the well--known correspondence between
atomistic lattices and T$_{_1}$--closure spaces (Faure, 1994), every ppl has
an associated closure space $(\Sigma, {\cal F}_{\cal L})$ where
$${\cal F}_{\cal L} = \{F \inn \Sigma\,|\, p\in \Sigma, p < \vee F
\Rightarrow p \in F\}$$ It easily follows that the above construction of the
minimal product generalizes to a minimal product of ppl's. To be precise,
the minimal product of $({\cal L}_1,\Sigma_1,\perp)$ and $({\cal
L}_2,\Sigma_2,\perp)$ is $({\cal L}_1\coprod{\cal L}_2,
\Sigma_1\times\Sigma_2,\perp)$, where
$(\Sigma_1\times\Sigma_2,{\cal L}_1\coprod{\cal L}_2)$ is the ${\bf
Cls}_{_1}$--product of $(\Sigma_1,{\cal F}_{{\cal L}_1})$ and
$(\Sigma_2,{\cal F}_{{\cal L}_2})$
and the orthogonality is defined as above.

We call $f:{\cal L}\rightarrow {\cal L}$ a {\em
symmetry} (of ppl's) if it is an order--automorphism, such that $\forall p,q
\in
\Sigma$ we have $p \perp q \Leftrightarrow f(p)\perp f(q)$. We remark that
for state spaces, symmetries are nothing else than permutations conserving
the orthogonality in both directions (Piron 1989). Indeed, if $\alpha$ is
such a permutation of $(\Sigma,\perp)$, then
$$f:{\cal L}_{\Sigma} \rightarrow {\cal L}_{\Sigma}: A \mapsto \alpha(A)$$
is the unique ortho--automorphism of ${\cal L}_{\Sigma}$ such that
$f\{p\}=\alpha(p)$ for every $p$ in $\Sigma$. In particular, $f$ is a
symmetry of the ppl $({\cal L}_{\Sigma},\Sigma,\perp)$ associated to
$(\Sigma,\perp)$.

We call a ppl $({\cal L},\Sigma,\perp)$  {\em plane transitive} if for
all atoms $p,q\in \Sigma$ there exist two distinct atoms $p_1, p_2$ and a
symmetry
$f$ such that $f|_{[0,p_1\vee p_2]}$ is the identity and $f(p)=q$. Looking
at Proposition~\ref{prop1}, it is obvious that if $\cal L$ is the
lattice of biorthogonal subspaces of an infinite--dimensional orthomodular
space $E$,
$E$ is a classical Hilbert space iff (with a slight abuse of language) $\cal
L$ is plane transitive.

\begin{proposition}
Let $({\cal L}_1,\Sigma_1,\perp)$ and $({\cal L}_2,\Sigma_2,\perp)$ be
ppl's. If both are plane transitive, then so is their minimal product $({\cal
L}_1\coprod{\cal L}_2,
\Sigma_1\times\Sigma_2,\perp)$.
\end{proposition}
Indeed, consider $(r_1,r_2)$ and $(s_1,s_2)$ in
$\Sigma_1\times\Sigma_2$. Choose a symmetry $f_1$ and an atom $p_1 \in
\Sigma_1$ such that
$f_1(r_1)=s_1$ and $f_1(p_1)=p_1$. Next,
choose
$p_2\neq q_2$ in
$\Sigma_2$ and a symmetry
$f_2$ of $({\cal L}_2,\Sigma_2,\perp)$ such that $f_2(r_2)=s_2$ and
$f_2|_{[0,p_2\vee q_2]}$ is the identical map. Then
$f_i|_{\Sigma_i}$ is a
${\bf Cls}_{_1}$--automorphism of $(\Sigma_i,{\cal F}_{{\cal L}_i})$. It
follows that $(t_1,t_2) \mapsto (f_1(t_1),f_2(t_2)) $  is a ${\bf
Cls}_{_1}$--automorphism of $(\Sigma_1\times\Sigma_2,{\cal L}_1\coprod{\cal
L}_2)$ and hence generates an order--automorphism $f_1\times f_2$ of ${\cal
L}_1\coprod{\cal L}_2$. Trivially, $f_1\times f_2 (r_1,r_2) = (s_1,s_2)$.
Also, $f_1\times f_2|_{[0,(p_1,p_2)\vee(p_1,q_2)]}$ is the identity.
Finally, it is straightforward to verify that $f_1\times f_2$ conserves
the orthogonality on $\Sigma_1\times\Sigma_2$ in both directions.

Using a similar argument, one easily shows the same holds for the
separated product. Note that a state space $(\Sigma,\perp)$ is called plane
transitive if its associated ppl $({\cal L}_{\Sigma},\Sigma,\perp)$ is plane
transitive.

\begin{proposition}
If two state spaces $(\Sigma_1,\perp)$ and $(\Sigma_2,\perp)$ are plane
transitive, then so is their separated product
$(\Sigma_1\times\Sigma_2,\perp_{_{\smallsep}})$.
\end{proposition}

\section{Questions}
\noindent
Several questions remain. Plane transitivity does not have the necessary
elegance to be a fundamental axiom: what is the physical significance of this
invariant plane? Another question is: can the unitary operators of an
orthomodular space be characterized at the lattice level? In other words,
can Holland's AUG axiom be formulated lattice theoretically? Maybe, it can be
generalized to the transitivity of the whole group of ortholattice
automorphisms and still characterize classical Hilbert spaces among
infinite--dimensional orthomodular spaces. This would be an elegant symmetry
(or evolution) axiom.

\section*{Acknowledgements}
\noindent
D.A. is a senior research associate and B.V.S. is a
research assistant of the Fund for Scientific Research---Flanders.

\section*{References}

\begin{description}

\item Aerts, D. (1982). Description of many physical entities without the
paradoxes encountered in quantum mechanics, {\em Found.\ Phys.}, {\bf 12},
1131-1170.

\item Aerts, D. (1983). Classical theories and non classical theories as a
special case of a more general theory, {\em J. Math.\ Phys.}, {\bf
24}, 2441-2454.

\item Aerts, D. (1984). Construction of the tensor product for the lattices
of properties of physical entities, {\em J. Math.\ Phys.}, {\bf 25},
1434-1441.

\item Aerts, D., Colebunders, E., Van der Voorde, A., and Van Steirteghem,
B. (1999). State property systems and closure spaces: a study of categorical
equivalence, {\em Int.\ J.\ Theor.\ Phys.}, {\bf 38}, 359-385.

\item Dikranjan, D., Giuli, E., and Tozzi, A. (1988). Topological categories
and closure operators, {\em Quaestiones Mathematicae}, {\bf 11}, 323-337.

\item Faure, Cl.-A. (1994). Categories of closure spaces and corresponding
lattices, {\em Cahier de top.\ et g\'eom.\ diff.\ cat\'eg.}, {\bf 35},
309-319.

\item Faure, Cl.-A., and Fr\"olicher, A. (1995). Dualities for
infinite--dimensional projective geometries, {\em Geom.\ Ded.}, {\bf 56},
225-236.

\item Holland, S.S. Jr.\ (1995). Orthomodularity in infinite dimensions; a
theorem of M. Sol\`er, {\em Bull.\ Amer.\ Math.\ Soc.}, {\bf 32}, 205-234.

\item Maeda,  F., and Maeda, S. (1970). {\em Theory of
symmetric lattices}, Springer--Verlag, Berlin.

\item Mayet, R. (1998). Some characterizations of the underlying division
ring of a Hilbert lattice by automorphisms, {\em Int.\ J.\ Theor.\ Phys.},
{\bf 37}, 109-114.

\item Moore, D.J. (1995). Categories of representations of physical
systems, {\em Helv.\ Phys.\ Acta}, {\bf 68}, 658-678.

\item Moore, D.J. (1999). On state spaces and property lattices, {\em
Stud.\ Hist.\ Phil.\ Mod.\
Phys.}, to appear.

\item Piron, C. (1976). {\em Foundations of quantum physics}, Benjamin,
New York.

\item Piron, C. (1989). Recent developments in quantum mechanics, {\em
Helv.\ Phys.\ Acta}, {\bf 62}, 82-90.

\item Piron, C. (1990). {\em M\'ecanique quantique bases et applications},
Presses polytechniques et universitaires romandes, Lausanne.

\item Pulmannov\`a, S. (1983). Coupling of quantum logics, {\em Int.\ J.\
Theor.\ Phys.}, {\bf
22}, 837.

\item Pulmannov\`a, S. (1985). Tensor product of quantum logics, {\em J.
Math.\ Phys.}, {\bf 26}, 1.

\item Sol\`er, M.P. (1995). Characterization of Hilbert spaces by
orthomodular spaces, {\em Comm.\ Algebra}, {\bf 23}, 219-243.
\end{description}

\end{document}